\documentclass[twocolumn, nofootinbib]{revtex4-1}
\usepackage{amsmath}
\usepackage{amsfonts}
\usepackage{amssymb}
\usepackage{epsfig}
\usepackage{graphicx}
\usepackage[dvipsnames]{xcolor}
\usepackage{xspace}
\usepackage{nicefrac}

\usepackage{tabu}
\usepackage{xfrac}
\usepackage{upgreek}
\usepackage{bbold}
\usepackage{ulem}

\usepackage{hyperref}
\usepackage{feynmf}

\usepackage{listings}

\newcommand{\be}{\begin{equation}}
\newcommand{\ee}{\end{equation}}
\newcommand{\bdm}{\begin{displaymath}}
\newcommand{\edm}{\end{displaymath}}
\newcommand{\bea}{\begin{eqnarray}}
\newcommand{\eea}{\end{eqnarray}}
\newcommand{\eq}[1]{Eq.~(\ref{#1})}
\newcommand{\eqs}[2]{Eqs.~(\ref{#1}) and (\ref{#2})}
\newcommand{\eqt}[2]{Eqs.~(\ref{#1})--(\ref{#2})}
\newcommand{\fig}[1]{Fig.~\ref{#1}}
\newcommand{\nn}{\nonumber}
\newcommand{\tab}[1]{Table~\ref{#1}}

\def\gsim{\raise0.3ex\hbox{$\;>$\kern-0.75em\raise-1.1ex\hbox{$\sim\;$}}}
\def\lsim{\raise0.3ex\hbox{$\;<$\kern-0.75em\raise-1.1ex\hbox{$\sim\;$}}}

\newcolumntype{C}{>{$}c<{$}}

\newcommand{\SARAH} {\texttt{SARAH}\xspace}
\newcommand{\SPheno} {\texttt{SPheno}\xspace}

\begin{document}
\preprint{KA-TP-21-2018, \today}

 \author{T. Faber}
 \email[E-mail: ]{thomas.faber@physik.uni-wuerzburg.de}
 \author{P. Meinzinger}
 \email[E-mail: ]{peter.meinzinger@stud-mail.uni-wuerzburg.de}
 \author{W. Porod}
 \email[E-mail: ]{porod@physik.uni-wuerzburg.de}
 \affiliation{
 Institut f\"ur Theoretische Physik und Astrophysik, Uni W\"urzburg, Germany
 }

 \author{M. Hudec}
 \email[E-mail: ]{hudec@ipnp.troja.mff.cuni.cz}
 \author{M. Malinsk\'y}
 \email[E-mail: ]{malinsky@ipnp.troja.mff.cuni.cz}
 \affiliation{
  Institute for Particle and Nuclear Physics, Charles University, Czech Republic
 }

 \author{F. Staub}
 \email[E-mail: ]{florian.staub@kit.edu}
\affiliation{Institute for Theoretical Physics (ITP), Karlsruhe Institute of Technology, Engesserstra{\ss}e 7, D-76128 Karlsruhe, Germany}
\affiliation{Institute for Nuclear Physics (IKP), Karlsruhe Institute of Technology, Hermann-von-Helmholtz-Platz 1, D-76344 Eggenstein-Leopoldshafen, Germany}

\title{A unified leptoquark model \\ confronted with lepton non-universality in $B$-meson decays
}

\hfill \parbox{5cm}{\vspace{ -1cm } \flushright KA-TP-21-2018}

\begin{abstract}
The anomalies in the $B$-meson sector, in particular $R_{K^{(*)}}$ and $R_{D^{(*)}}$, are often interpreted as hints for physics beyond the Standard Model. To this end, leptoquarks or a heavy $Z'$ represent the most popular SM extensions which can explain the observations. However, adding these fields by hand is not 
very satisfactory as it does not address the big questions like a possible embedding into a unified gauge theory. 
On the other hand, light leptoquarks within a unified framework are challenging 
due to additional constraints such as lepton flavor violation. The existing accounts typically deal with this issue by providing estimates on the relevant
couplings. In this letter we consider a complete model based on the $SU(4)_{\rm C}\otimes SU(2)_{\rm L}\otimes U(1)_{\rm R}$ gauge symmetry, a subgroup of $SO(10)$, featuring both scalar and vector leptoquarks.
We demonstrate that this setup has, in principle, all the potential to accommodate $R_{K^{(*)}}$ and $R_{D^{(*)}}$ while
respecting bounds from other sectors usually checked in this context. However, it turns out that 
$K_L \to e^{\pm} \mu^{\mp}$ severely constraints not only the vector but also the scalar leptoquarks and, consequently,
also the room for any sizeable deviations of $R_{K^{(*)}}$ from 1.
We briefly comment on the options for extending the model
in order to conform this constraint. Moreover, we present a simple criterion for all-orders proton stability within this class of models.
\end{abstract}

\maketitle

\section{Introduction}
In recent years a few anomalies in the B-meson sector have been observed by different
experiments. The most striking one is a 3.5-$\sigma$ deviation in the ratios
\begin{equation*}
R_{D^{(*)}} = \frac{\Gamma(\bar B \to D^{(*)} \tau \bar \nu)}
                   {\Gamma(\bar B \to D^{(*)} l \bar \nu)}
      \hspace{1cm}(l=e,\mu)        \,,     
\end{equation*}
with
\begin{equation}
R^\mathrm{exp}_D = 0.388 \pm 0.047 ,\quad  R^\mathrm{exp}_{D^*} = 0.321 \pm 0.021\,, 
\end{equation}
from the Standard Model (SM) lepton universality expectations  
\begin{equation}
 R^\mathrm{SM}_D = 0.300 \pm 0.010 ,\quad  R^\mathrm{SM}_{D^*} = 0.252 \pm 0.005 \,.
\label{eq:R_D}
\end{equation}
This was first reported by BaBar
\cite{Lees:2012xj,Lees:2013uzd}  consistent with
measurements by Belle \cite{Matyja:2007kt,Bozek:2010xy,Huschle:2015rga}.
Recently this has been confirmed by LHCb in case of
$R_{D^{*}}$ \cite{Aaij:2015yra} at the 2.1-$\sigma$ level. Additional deviations
from lepton universality have recently been reported
by LHCb in the ratio
\begin{eqnarray}
R_K  &=& \frac{\Gamma(\bar B \to \bar K \mu^+ \mu^-)}
                   {\Gamma(\bar B \to \bar Ke^+ e^-)}= 0.745^{+0.090}_{-0.074} \pm 0.036 \,\,,\,\, \\
R_{K^*}  &=&                    
\label{eq:R_K} \frac{\Gamma(\bar B \to \bar K^* \mu^+ \mu^-)}
                   {\Gamma(\bar B \to \bar K^* e^+ e^-)}= 0.69^{+0.11}_{-0.07} \pm 0.05
\end{eqnarray}
in the dilepton invariant mass bin 1 GeV$^2 \le q^2 \le 6$ GeV$^2$
\cite{Aaij:2014ora,Aaij:2017vbb}. 
These ratios are predicted to be 1 within the SM and are practically free from
theoretical uncertainties.
Equally intriguing is a discrepancy in
the angular observables in the rare 
$\bar B \to \bar K \mu^+ \mu^-$ decays measured by LHCb
\cite{Aaij:2013qta} which, however, is subject to significant
hadronic uncertainties \cite{Beaujean:2013soa,Lyon:2014hpa}. While the individual
discrepancies are between 2 and 3 $\sigma$, they all point in the same direction
and amount to more than 4.5-$\sigma$ deviations once combined in a fit 
\cite{Altmannshofer:2017fio,Capdevila:2017bsm}. 

In Refs.~\cite{Bauer:2015knc,Bauer:2015boy} it has been shown that the deviations in $R_D$ and $R_K$ can
be explained by an effective model adding one generation of scalar leptoquarks (LQs)
with the quantum numbers of the right-handed $d$-quark and an additional scalar gauge singlet
which couples to the LQs. However, it has been shown that this leads to a too large rate
for $b \to s \nu \nu$  \cite{Crivellin:2017zlb}. In Ref.~\cite{Chao:2015nac} another model with
two different LQs, one with gauge quantum numbers of the right-handed $d$-quark
and one with charge $4/3$, has been presented which explains also
 neutrino masses at the 2-loop level. As has been shown in Refs.~\cite{Murphy:2015kag,Alonso:2015sja,Calibbi:2015kma,Fajfer:2015ycq,Hiller:2016kry,Bhattacharya:2016mcc,Buttazzo:2017ixm,Kumar:2018kmr}, another possibility to successfully
accommodate the data is to use  vector
LQs.
A somewhat more complete model containing two types of
vector LQs to explain the two-photon excess,  based on a Froggatt-Nielsen ansatz for the required coupling structures, has been presented in~\cite{Deppisch:2016qqd}. 
Beside the above mentioned violations of lepton-universality this model is also compatible with the
neutrino data. Another possibility is that the required leptoquarks are bound states
of strongly interacting fermions~\cite{Barbieri:2017tuq}.

Most of these settings are effective models containing just the  pieces required to account for
the discussed experimental observations, which is clearly the first
logical step to make when a new signal shows up.
However, eventually one would like to understand the observations from a more fundamental
perspective. Several attempts
in this direction exist already in the literature \cite{Barbieri:2015yvd,Barbieri:2016las,Assad:2017iib,DiLuzio:2017vat,Calibbi:2017qbu,%
Bordone:2017bld,Barbieri:2017tuq,Dorsner:2017ufx,Blanke:2018sro,Greljo:2018tuh,Bordone:2018nbg,Matsuzaki:2018jui}.
Of course, the most attractive scenario would be a UV completion
compatible with theoretical requirements like gauge-coupling unification with the
potential to explain also the observed dark matter relic density.

From the GUT perspective the Pati-Salam (PS) model \cite{Pati:1974yy}
emerges as the first and very natural candidate for a low-energy gauge framework featuring vector as well as 
scalar leptoquarks within a simple dynamical and renormalizable scheme. However, the Kibble-Zurek mechanism of the early-Universe monopole creation~\cite{Zurek:1985qw} suggests that the PS-breaking should occur above the inflation scale~\cite{Murayama:2009nj}. It is therefore advisable to choose 
$SU(4)_\mathrm C\otimes SU(2)_\mathrm L\otimes U(1)_\mathrm R$ instead as a gauge group of a potentially viable model (as in Refs.~\cite{Smirnov:1995jq,Perez:2013osa}) which, indeed, does not suffer from the monopole issue.

The structure of this letter is as follows: in Section \ref{sec:model} we present the model
and discuss the possibilities to obtain leptoquarks with masses in the TeV range. 
In Section \ref{sec:low_energy} we discuss in which parts of the parameter space the $B$ 
anomalies could be accounted for and what are the constraints from the existing 
low energy data. In Section \ref{sec:conclusions} we draw our conclusions. 
In Appendix \ref{sec:proton_decay} we demonstrate
that in this class of models proton 
remains stable to all orders in perturbation theory.

\section{Model description}
\label{sec:model}
In what follows we consider the model proposed by Fileviez-Perez and Wise in Ref. \cite{Perez:2013osa}. 
For convenience, we briefly outline it here, focusing on the features related to flavor physics.

The model is based on the gauge group $G=SU(4)_\mathrm{C}\otimes SU(2)_\mathrm{L}\otimes U(1)_\mathrm{R}$, 
where the first factor unifies the three colors of quarks with the 
lepton number. 
This group is spontaneously broken to $G_\mathrm{SM}= SU(3)_\mathrm{c}\otimes SU(2)_\mathrm{L} \otimes U(1)_\mathrm{Y}$ and further down to $G_\mathrm{vac}=SU(3)_\mathrm{c}\otimes U(1)_\mathrm{Q}$, following the branching rules
\begin{equation}
Y=R+\sfrac{1}{2}\,[B-L]\,,\qquad Q=T_\mathrm{L}^{3}+Y,
\end{equation}
where\footnote{We use the square brackets here in order to indicate an indivisible symbol.}
\begin{equation}
\label{T15}
[B-L] =\sqrt{8/3} \;T^{15}_\mathrm{C}= \mathrm{diag}\left(+\tfrac{1}{3},+\tfrac{1}{3},+\tfrac{1}{3},-1\right)\,.
\end{equation}
The matching condition for the QCD coupling at the scale, where $SU(4)_{\rm C}$ is broken, is simply $g_3=g_4$. 

The entire field content of the model is summarised in \tab{tab:U1charges}. We also include information about other $U(1)$ charges which we need in Appendix \ref{sec:proton_decay} where the details of the baryon number conservation and lepton number violation are discussed.

\begin{table*}
\begin{tabular}{|C||C|C|C||C|C|C|C|C|} 
\hline 
	& G  & G_\mathrm{SM} & G_\mathrm{vac} &  [B\!-\!L]  &  F  &   M  &  B  &  L, L'
\\ \hline \hline
\multicolumn{9}{|c|}{\bf Fermions }\\
\hline
F_ L=\begin{pmatrix}  Q\\L \end{pmatrix} 
& (4,2,0) & \begin{array}{cc} Q  & (3,2,\sfrac{1}{6}) \\ \\ L & (1,2,\sfrac{-1}{2})  \end{array} & 
\begin{array}{cc} u & (3,\sfrac{2}{3}) \\ d & (3,\sfrac{-1}{3}) \\ \nu & (1,0) \\ e & (1,-1)  \end{array} &
   \begin{pmatrix} \sfrac{+1}{3} \\-1  \end{pmatrix} & 
   +1 & +1 & 
   \begin{pmatrix} \sfrac{+1}{3} \\0  \end{pmatrix} &  
   \begin{pmatrix} 0 \\ +1   \end{pmatrix} 
\\ \hline 
f_{u}^c = \begin{pmatrix}u^c & \nu^c \end{pmatrix}
& (\overline{4},1,\sfrac{-1}{2})
& \begin{array}{cc} u^{c}  & (\overline{3},1,\sfrac{-2}{3}) \\ \nu^{c} & (1,1,0) \end{array}  & \begin{array}{c} (\overline{3},\sfrac{-2}{3}) \\ (1,0)  \end{array}  & 
	\begin{pmatrix} \sfrac{-1}{3} & 1  \end{pmatrix} &
	-1 &  -1 &
	\begin{pmatrix} \sfrac{-1}{3}  & 0 \end{pmatrix} &
	\begin{pmatrix} 0  & -1   \end{pmatrix} 
\\ \hline
f_{d}^c = \begin{pmatrix} d^c & e^c  \end{pmatrix} & (\overline{4},1,\sfrac{1}{2})
& \begin{array}{cc} d^{c} & (\overline{3},1,\sfrac{1}{3}) \\ e^{c} & (1,1,1) \end{array}  & \begin{array}{c} (\overline{3},\sfrac{1}{3}) \\ (1,1) \end{array}  & 
	\begin{pmatrix} \sfrac{-1}{3} & 1  \end{pmatrix} &
	-1 & -1 &
	\begin{pmatrix} \sfrac{-1}{3}  & 0 \end{pmatrix} &
	\begin{pmatrix} 0  & -1   \end{pmatrix} 
\\ \hline
N & (1,1,0)  & (1,1,0)  & (1,0)  &0&+1&0&0&0,+1
\\ \hline \hline
\multicolumn{9}{|c|}{\bf Scalars}\\
\hline
\chi= \begin{pmatrix} {\bar S_1}^\dagger \\ \chi^0  \end{pmatrix} & (4,1,\sfrac{1}{2}) 
& \begin{array}{cc} {\bar S_1}^\dagger & (3,1,\sfrac{2}{3}) \\ \chi^0 & (1,1,0) \end{array}  & \begin{array}{c} (3,\sfrac{2}{3}) \\ (1,0) \end{array}  & 
	\begin{pmatrix} \sfrac{+1}{3} \\ -1 \end{pmatrix} &
	0 & +1 &
	\begin{pmatrix}  \sfrac{+1}{3} \\ 0 \end{pmatrix} &
	\begin{pmatrix} 0 \\ +1 \end{pmatrix},
	\begin{pmatrix} -1 \\ 0 \end{pmatrix}  
\\ \hline
H & (1,2,\sfrac{1}{2}) & (1,2,\sfrac{1}{2})  & \begin{array}{cc} H_1^+ & (1,1) \\ H_1^0 & (1,0) \end{array}  & 0&0&0&0&0 
\\ \hline
\begin{array}{r}
\Phi = \begin{pmatrix} G & R_2 \\ \tilde{R}_2^\dagger & 0  \end{pmatrix}
\\\\+ \sqrt{2}\, T^{15} H_2 
\end{array}
& (15,2,\sfrac12) &\begin{array}{cc}
                    R_2 & (3,2,\sfrac76)\\ \\
                    \tilde{R}_2^\dagger & (\overline{3},2,\sfrac{-1}{6}) \\ \\
                    G & (8,2,\sfrac12) \\ \\
                    H_2 & (1,2,\sfrac12)
                   \end{array}
  & \begin{array}{cc}
                    R^{5/3}_2 & (3,\sfrac53)\\
                    R^{2/3}_2 & (3,\sfrac23)\\
                    \tilde{R}_2^{-1/3\,\dagger} & (\overline{3},\sfrac13) \\
                                        \tilde{R}_2^{2/3\,\dagger} & (\overline{3},\sfrac{-2}{3}) \\
                    G^+ & (8,1) \\
                    G^0 & (8,0) \\
                    H^+_2 & (1,1) \\
                    H^0_2 & (1,0) 
                   \end{array} & 
	\begin{pmatrix} 0 &\sfrac{+4}{3} \\ \sfrac{-4}{3} & 0  \end{pmatrix} &
	0&0&
	\begin{pmatrix} 0 &\sfrac{+1}{3} \\ \sfrac{-1}{3} & 0  \end{pmatrix} &
	\begin{pmatrix} 0&-1 \\ +1&0  \end{pmatrix} 
\\ \hline \hline
\multicolumn{9}{|c|}{\bf Gauge Bosons}\\
\hline
\begin{array}{c} A_\mu =\begin{pmatrix} G_\mu & X_\mu \\ X^*_\mu & 0  \end{pmatrix}  \\ +  T^{15} B'_\mu \end{array} & (15,1,0)
& \begin{array}{cc} G_\mu & (8,1,0) \\ X_\mu & (3,1,\sfrac23) \\ B'_\mu & (1,1,0) \end{array}  &  \begin{array}{c} (8,0) \\ (3,\sfrac23)   \\ (1,0) \end{array} & 
	\begin{pmatrix} 0 &\sfrac{+4}{3} \\ \sfrac{-4}{3} & 0  \end{pmatrix} &
	0&0&
	\begin{pmatrix} 0 &\sfrac{+1}{3} \\ \sfrac{-1}{3} & 0  \end{pmatrix} &
	\begin{pmatrix} 0&-1 \\ +1&0  \end{pmatrix} 
\\ \hline
W_\mu  & (1,3,0) & (1,3,0)  & \begin{array}{cc} W_\mu^\pm & (1,\pm1) \\ W_\mu^3 & (1,0) \end{array}  & 0&0&0&0&0
\\
\hline
 B_\mu  & (1,1,0)  & (1,1,0)  & (1,0)  & 0&0&0&0&0
\\ \hline
\end{tabular}
\caption{The field content of the model together with all gauge quantum numbers for the different regimes as well as charges under several other global $U(1)$'s. Whenever $L=L'$ only the common value is displayed.}
\label{tab:U1charges}
\end{table*}

The SM fermions together with the right-handed neutrinos are combined into three quadruplets under $SU(4)_{\rm C}$ appearing in 
three copies representing different generations. On top of that, three 
fermionic gauge singlets $N$ necessary to generate the 
correct neutrino masses via inverse seesaw~\cite{Mohapatra:1986bd} are added. 

The gauge field sector corresponding to $SU(4)_{\rm C}$ consists of the gluons, $Z'$ and a vector leptoquark 
$X\sim (3,1,+\tfrac{2}{3})$ which mediates flavor violating processes such as $K_L\rightarrow e^{\pm} \mu^{\mp}$. 
This tight constraint implies that, for standard-size couplings, the mass of the vector leptoquark has to be at least of the order of 1.6$\times 10^3$  TeV~\cite{Valencia:1994cj}\footnote{Note that this limit may be significantly reduced if the freedom in the mixing in the charged leptoquark currents is fully exploited, see, e.g.,~\cite{Smirnov:2018ske} and references therein.}. 

The scalar sector consists of three multiplets $\chi, \Phi$ and $H$, see \tab{tab:U1charges}. The most general renormalizable scalar potential  
for these fields reads
\begin{align}
&V= m_H^2 |H|^2 + m_\chi^2 |\chi|^2 + m_\Phi^2 {\rm{Tr}} (|\Phi|^2) + \lambda_1 |H|^2 |\chi|^2 \nn\\
& + \lambda_2 |H|^2  {\rm{Tr}} (|\Phi|^2) 
+ \lambda_3 |\chi|^2  {\rm{Tr}} (|\Phi|^2) 
+ ( \lambda_4 H^\dagger_i \chi^\dagger \Phi^i \chi + {\rm{h.c.}} ) 
\nn\\&
+ \lambda_5 H^\dagger_i {\rm{Tr}} (\Phi^\dagger_j \Phi^i)  H^j 
+ \lambda_6 \chi^\dagger \Phi^i \Phi^\dagger_i \chi
+ \lambda_7 |H|^4 + \lambda_8 |\chi|^4
\nn\\&
+ \lambda_9 {\rm{Tr}} (|\Phi|^4)   
+  \lambda_{10} ({\rm{Tr}} |\Phi|^2)^2
+\Big( \lambda_{11} H^\dagger_i \;\mathrm{Tr}( \Phi^{i} \, \Phi^{j})  H^\dagger_j
\nn\\ &
+\lambda_{12} H^\dagger_i \;\mathrm{Tr}(\Phi^{i}\, \Phi^{j} \,\Phi^{\dagger }_{j})
+
\lambda_{13} H^\dagger_i \;\mathrm{Tr}( \Phi^{i} \,\Phi^{\dagger }_{j} \,  \Phi^{j})  + \mathrm{h.c.}\Big) \nn\\
& 
+\lambda_{14} \chi^\dagger  |\Phi|^2  \chi
+\lambda_{15} \mathrm{Tr}( \Phi^{\dagger}_{i}\, \Phi^{j}_{}\, \Phi^{\dagger}_{j}\, \Phi^{i}_{})\nn \\
& 
+\lambda_{16} \mathrm{Tr}( \Phi^{\dagger}_{i}\, \Phi^{j}_{})\,
\mathrm{Tr}( \Phi^{\dagger}_{j}\, \Phi^{i}_{})
+\lambda_{17} \mathrm{Tr}( \Phi^{\dagger}_{i}\, \Phi^{\dagger}_{j})\, \mathrm{Tr}( \Phi^{i}_{}\, \Phi^{j}_{}) \nn\\
& +\lambda_{18} \mathrm{Tr}( \Phi^{\dagger}_{i}\, \Phi^{\dagger}_{j}\, \Phi^{i}_{}\, \Phi^{j}_{})
+\lambda_{19} \mathrm{Tr}( \Phi^{\dagger}_{i}\, \Phi^{\dagger}_{j}\, \Phi^{j}_{}\, \Phi^{i}_{})
\label{eq:Vscalar}
\end{align}
with $|H|^2=H^\dagger_i H^i$, $|\chi|^2=\chi^\dagger \chi$, $|\Phi|^2=\Phi^\dagger_i \Phi^i$ 
where $i,j$ are the $SU(2)$ indices. The trace is taken over the $SU(4)_{\rm C}$ indices only.
Notice that the terms proportional to $\lambda_{11}, \ldots,\lambda_{19}$ have been omitted in the original paper 
\cite{Perez:2013osa}; we include them here for completeness.

The breaking of the $SU(4)_{\rm C}$ group as well as the electroweak symmetry breaking is triggered by the corresponding 
vacuum expectation values\footnote{The round and square brackets are used to distinguish between the $SU(4)_{\rm C}$ and $SU(2)_{\rm L}$ multiplets, respectively.} (VEVs)
\begin{align}
\begin{split}
\langle \chi \rangle 
&= \frac{1}{\sqrt{2}}\begin{pmatrix} 0 \\ v_\chi \end{pmatrix},
\qquad 
\langle H \rangle = \frac{1}{\sqrt{2}}\begin{bmatrix}  0 \\ v_1 \end{bmatrix},
\\
\langle \Phi \rangle &= \frac{1}{2\sqrt{6}} 
\begin{pmatrix}\mathbb{1} & 0 \\ 0 &-3\end{pmatrix} 
\otimes \begin{bmatrix}0\\v_2\end{bmatrix},
\end{split}
\end{align}
which are parametrised by $v_1=v_\mathrm{ew} \sin\beta$, $v_2=v_\mathrm{ew} \cos\beta$,  
with $v_\mathrm{ew} \simeq 246$~GeV. The SM-like Higgs $h$ is a superposition of the fields Re($\{H_1^0,H_2^0,\chi^0\}$). 

The fermion masses are generated by the following interactions between the scalars and fermions:
\begin{align}
\label{YukawaL}
\begin{split}
- \mathcal{L}_Y   &= f^c_u Y_1 H F_L + f^c_u Y_2 \Phi F_L + f^c_d Y_3 H^\dagger F_L \\
& +f^c_d Y_4 \Phi^\dagger F_L + f^c_u Y_5 \chi N + \frac12  N \mu N + \text{h.c.}
\end{split}
\end{align}
In the broken phase, this leads to the following relations between the mass matrices for the SM fermions and the underlying Yukawa matrices:
\begin{align}
U_u^\dagger \hat{M}_u V_u = &\frac{v_1}{\sqrt{2}} Y_1 + \frac{v_2}{2 \sqrt{6}} Y_2, \\
\label{eq:massD}
U_d^\dagger \hat{M}_d V_d = &\frac{v_1}{\sqrt{2}} Y_3 + \frac{v_2}{2 \sqrt{6}} Y_4, \\
\label{eq:massL}
\hat{M}_e= & \frac{v_1}{\sqrt{2}} Y_3 - \frac{3 v_2}{2\sqrt{6}} Y_4.
\end{align}
Here $\hat{M}$ are diagonal and $U$, $V$ are unitary matrices describing the relation between the gauge and mass eigenstates. We work in a basis where the lepton mass matrix is flavor-diagonal.
The only constraints on 
$U$'s and $V$'s are that $V_{\rm CKM} = V_u^\dagger V_d$ must be reproduced. 
In the current study, we shall assume that all the Yukawa matrices in~(\ref{YukawaL}) are symmetric in the flavor space and, hence, $U_d=V_d^*$ and 
$U_u=V_u^*$; besides simplicity, this is motivated by the idea that this model might eventually be embedded in a variant of the minimal $SO(10)$ framework
(such as proposed in Ref.~\cite{Bertolini:2012im}).
Thus, we are left with just one  mixing matrix $V_d$ which we can choose freely. As we will see, this freedom
is crucial for accommodating the $B$ anomalies without violating other constraints from lepton flavor violating observables.

In the neutrino sector we have a $9\times 9$ complex symmetric matrix in the basis $(\nu,{\nu}^c,N)$ 
which should yield the  light-neutrino PMNS matrix as well as their measured mass differences.  

The scalar leptoquarks in this model ($R_{2}$, $\tilde{R}_{2}$ and $S_{1}$) reside in  $\Phi$ and $\chi$. 
After the $SU(4)_{\rm C}$ breaking, the masses of the $SU(2)$-doublet scalars conform the sum-rule
\be
m^{2}_{G}+2 m^{2}_{H'}\sin^{2}\beta=\frac{3}{2}(m^{2}_{R_{2}}+m^{2}_{\tilde{R}_{2}})\,,
\ee
where $G$ denotes the scalar
gluons and $H'$ stands for the heavier eigenstate of the $H$ and $H_{2}$ mixture.
This, among other things, implies that one can not have both  $m^{2}_{R_{2}}$ and $m^{2}_{\tilde{R}_{2}}$ significantly smaller than $m^{2}_{G}$ and $ m^{2}_{H'}$.
It is well known~\cite{DAmico:2017mtc}, however, that a light scalar leptoquark with the quantum numbers of $R_{2}$  is way more suitable for a proper explanation of $R_{K^{(*)}}$ than $\tilde{R}_{2}$; hence, in what follows, we shall work in the setting with $m^{2}_{R_{2}}\ll m^{2}_{\tilde{R}_{2}}$. 

Furthermore,  the mixing among the charge-2/3 components of  $R_2$ and $\tilde{R}_2$ with the ${\bar S_1}$ field emerges only from the $SU(2)_{\rm L}\otimes U(1)_{\rm Y}$ breaking. Hence, the physical mass eigenstates $R'_{1,2}$ are dominated by  $R_2^{2/3}$ and $\tilde{R}_2^{2/3}$, respectively, whereas ${\bar S_1}$ approximates the Goldstone mode associated to the vector leptoquark.

%



\section{Low energy observables}
\label{sec:low_energy}
\begin{figure}[tb]
\includegraphics[width=0.5\linewidth]{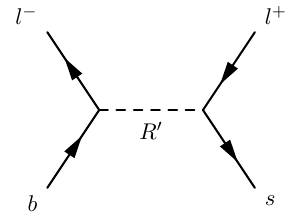} 
\caption{Tree-level contribution to $R_K$ via leptoquarks $R'$.}
\label{fig:DiaRK}
\end{figure}
\begin{figure}[tb]
\includegraphics[width=0.75\linewidth]{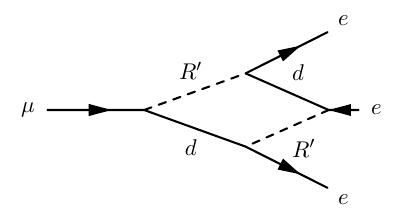} 
\caption{One-loop contribution to $\mu\to 3e$ via leptoquarks $R'$.}
\label{fig:DiaMu3e}
\end{figure}

We turn now to a discussion of the relevant low-energy observables in the current model. First of all, we want 
to explain the observed deviation from the lepton universality in the $B$-meson decays. The Feynman diagram responsible 
for the tree-level contributions to $R_K$ via the scalar leptoquarks $R'$ is depicted in \fig{fig:DiaRK}. It is important to notice, however, that
the same leptoquark which should explain the $B$-meson anomalies would also contribute to other observables. 
At the tree level, one can expect an impact on other meson observables like $B\to X_s \nu\nu$, $B_s^0\to l\bar{l}$ or $K_L \to e\mu$. Moreover, there are important loop  contributions to $B\to X_s \gamma$, $l\to l' \gamma$ and $l\to 3 l'$.
An example of the responsible Feynman diagrams is shown in \fig{fig:DiaMu3e}.  

For our numerical study we used the {\tt Mathematica} package \SARAH \cite{Staub:2008uz,Staub:2009bi,Staub:2010jh,Staub:2012pb,Staub:2013tta} 
and extended\footnote{Details about the new feature to support 
unbroken subgroups in \SARAH will be given in \cite{longpaper}.} it to support the model under consideration.
In the first step, we used the model files to produce a spectrum generator based on 
\SPheno~\cite{Porod:2003um,Porod:2011nf}. \SPheno calculates the mass spectrum providing the 
option 
to include all one-loop and the important two-loop corrections to neutral scalar masses 
\cite{Goodsell:2014bna,Goodsell:2015ira,Braathen:2017izn} in the DR or the MS scheme. However, we 
are assuming here a 
full on-shell calculation of all masses, i.e., all shifts can be absorbed into counter-terms of the couplings leaving the mass spectrum unchanged. 

In addition, \SPheno provides an interface to  {\tt HiggsBounds} \cite{Bechtle:2008jh,Bechtle:2011sb,Bechtle:2013wla} 
which we used to check the constraints on the neutral scalars. Moreover, \SPheno 
calculates electroweak precision  as well as flavor observables. The calculation of flavor 
observables is based on the {\tt FlavorKit} functionality presented in 
Ref.~\cite{Porod:2014xia}. We used this 
feature in order to calculate the values for all necessary lepton flavor violating 
observables including $K_L\to e \mu$. Moreover, the values of the Wilson coefficients relevant for the
 $B$-physics 
calculated by \SPheno were passed to {\tt flavio} 
\cite{david_straub_2017_897989} to obtain predictions for the $B$-meson observables.

\begin{table}
\begin{tabular}{|c|c|}
\hline
\multicolumn{2}{|c|}{Numerical input values}     \\
\hline
\hline
\rule{0pt}{10pt}$Y_2$ & $ \text{diag}(10^{-8}, 10^{-7}, 10^{-5}) $ \\
\hline
\rule{0pt}{10pt}$Y_5$ & $\text{diag} (10^{-2}, 5 \cdot 10^{-2}, 10^{-1}) $ \\
\hline
\rule{0pt}{10pt}$\theta_{12}, \ \theta_{13}, \ \theta_{23}$ & $ \nicefrac{\pi}{2}, \ 0,\ \nicefrac{\pi}{4} $ \\
\hline
\rule{0pt}{10pt}$v_\chi$ & $ 4 \cdot 10^6 $~GeV  \\
\hline
\rule{0pt}{10pt}$m_A,\ m_{R'_1} $ & $ 2 \cdot 10^5$~GeV, \ 900~GeV  \\
\hline
\rule{0pt}{10pt}$\tan \beta$ & 50  \\
\hline
\end{tabular}
\caption{Summary of the default input values used in this analysis except if stated otherwise. All other BSM scalars have masses of the order $\mathcal{O}(m_A)$.}
\label{tab:parameters}
\end{table}

In \tab{tab:parameters} we collect the input parameters for this study. The remaining parameters
affect the heavy states which do not contribute to the observables discussed below. For the fermions
we take as input
the known quark and lepton masses, the CKM and the PMNS matrices using the best fit values
reported in \cite{PDG2018}, the Yukawa couplings $Y_2$, $Y_5$
and $\tan\beta$.
For an explanation of the $B$-physics observables we need an off-diagonal structure in $Y_4$. We therefore
take $V_d$ as input, parametrizing it as
\begin{align}
V_d  = &
\begin{pmatrix}
1 && 0 && 0 \\ 
0 && c_{23} && s_{23} \\ 
0  && -s_{23} && c_{23}
\end{pmatrix}
\cdot 
\begin{pmatrix}
c_{13} && 0 && s_{13} e^{i \delta} \\ 
0 && 1 && 0 \\ 
-s_{13} e^{i \delta}  && 0 && c_{13}
\end{pmatrix}
\notag 
\\
\cdot & 
\begin{pmatrix}
c_{12} && s_{12} && 0 \\ 
-s_{12} && c_{12} && 0 \\ 
0  && 0 && 1
\end{pmatrix}\,,
\end{align} 
denoting $c_{ij}=\cos\phi_{ij}$ and $s_{ij}=\sin\phi_{ij}$, and
vary all three angles in the range $[0,\pi]$ with $\delta=0$ for simplicity. 
 In the scalar sector we take the $SU(4)_{\rm C}$-breaking VEV $v_\chi$, the mass of the $R'_1$ leptoquark and the overall scale ($m_{A}$) of the heavy integer-charge Higgs bosons as input.
 Moreover,
we fix the mass of the SM-like Higgs-boson to $125.1$~GeV by adjusting $\lambda_7$ accordingly. As we take the
heavy Higgs bosons to be in the multi-TeV range, we are in the decoupling limit and fulfil automatically the
experimental constraints on the observed Higgs boson.

As long as $R_{D}$ and $R_{D^{*}}$ are concerned, there are in principle two ways to accommodate the data in the current scenario:  (i) The model automatically contains a vector leptoquark of a suitable type~\cite{Calibbi:2017qbu}; however, its potential effects in $R_{D^{(*)}}$ are strongly suppressed by the need to be compatible with $K_{L}\to \mu e$ which pushes its allowed mass above 1600 TeV. (ii) A way larger effect than (i) is expected if the two scalar $SU(2)$-doublet leptoquarks in the spectrum remain light and, at the same time, their charge $+2/3$ components entertain a large mixing. This, as discussed in Sect.~\ref{sec:model}, is impossible in the current setting. Hence, the simple model at stakes has serious issues with accommodating the existing $\overline{B}\to D^{(*)}\ell \bar\nu$ data. 

\begin{figure}[t]
\includegraphics[scale=0.75]{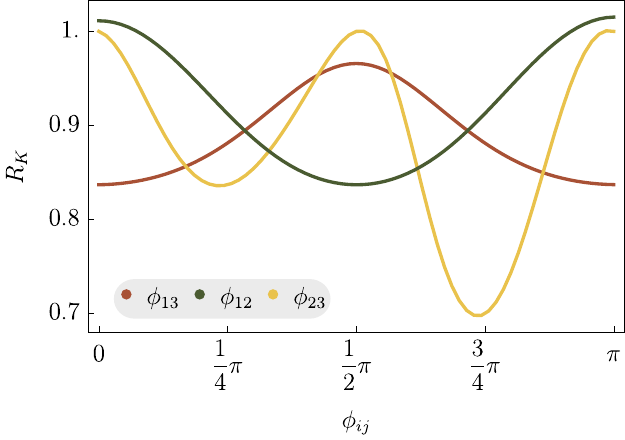} \\[5mm]
\includegraphics[scale=0.75]{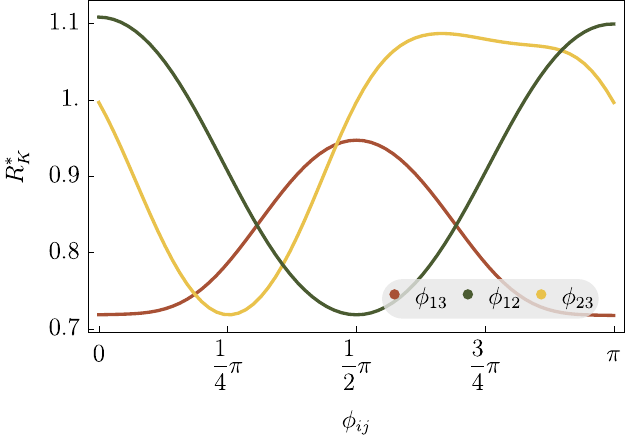} 
\caption{$R_K$ (first row) and $R_{K^*}$ (second row) as a function of the mixing angles
in the down-quark sector. The remaining parameters are given in \tab{tab:parameters}. }
\label{fig:RKangles}
\end{figure}

By contrast, it is fairly easy to explain the currently observed values for $R_K$ and $R_{K^*}$ as demonstrated
in \fig{fig:RKangles}. Here we have fixed the input parameters as given in \tab{tab:parameters} and varied one of the
relevant angles. The new physics contributions to the Wilson coefficients are given by
\begin{align}
C_{9,R_2}^{\mathrm{NP}\,ij} = C_{10,R_2}^{\mathrm{NP}\,ij} = - \frac{1}{4} \frac{1}{m_{R'_1}^2} \left( Y_4 V_d^\dagger \right)_{i3} \left(V_d  Y_4^\dagger \right)_{2j}\,, 
\end{align}
where $i = j = 1, 2$ for the contribution to $b \rightarrow s e e$ and $b \rightarrow s \mu \mu$, respectively. 
 Using \eqs{eq:massD}{eq:massL} we calculate the Yukawa couplings $Y_3$ and $Y_4$ in terms
of the lepton and quark masses as well as $V_d$. Exploiting
the hierarchy in the fermion mass spectrum we get, to a good approximation,
\begin{align}
\label{eq:C9R2}
C_{9,R_2}^{\mathrm{NP} ee} \simeq & - \frac{3 m_b m_s}{128 m^2_{R'_1} v_{\rm ew}^2} (1+\tan^2\beta) f_-\,,\\
C_{9,R_2}^{\mathrm{NP} \mu\mu} \simeq &  \frac{3 m_b (m_s - m_\mu)}{128 m^2_{R'_1} v_{\rm ew}^2} (1+\tan^2\beta) f_+\,,
\label{eq:C10R2}
\end{align}
with 
\begin{align}
f_{\pm} &= 4 \cos2\phi_{23} \ \sin2\phi_{12} \ \sin\phi_{13} \nonumber \\
&+ \sin2\phi_{23}\left(3 \cos2\phi_{12} \pm 1 \pm 2 \cos2\phi_{13}\ {}^{\sin^{2}\phi_{12}}_{\cos^{2}\phi_{12}}\right).
\end{align}
Combining this with
\begin{equation}
R_K \simeq 1 + \frac{1}{|C_9^{\mathrm{SM}}|^2} \left(|C_{9,R_2}^{\mathrm{NP} \mu\mu}|^2-|C_{9,R_2}^{\mathrm{NP} ee}|^2\right)
\end{equation}
from  \cite{Hiller:2014ula} one obtains an excellent analytic approximation to the
numerical results shown in \fig{fig:RKangles}.

\begin{figure}[t]
\includegraphics[width=\linewidth]{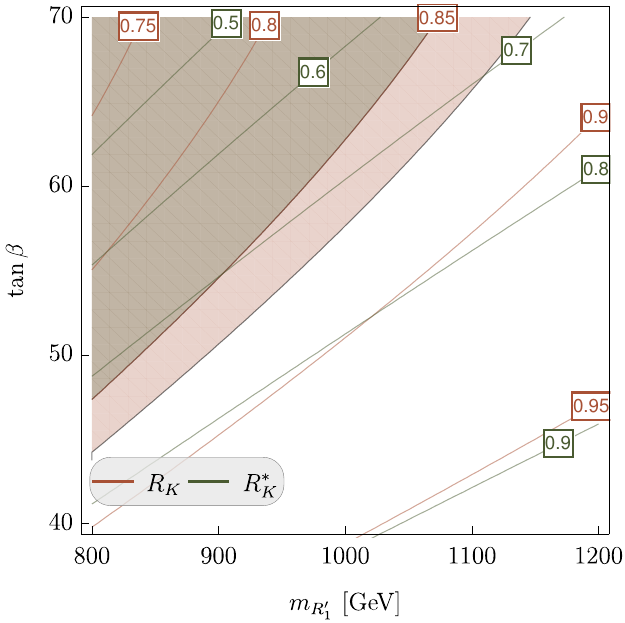}
\caption{Contour plots of $R_K$ and $R_{K^*}$  as a function of the leptoquark mass
and $\tan\beta$. The shaded regions indicate the range preferred by current data. 
The remaining parameters are given in 
\tab{tab:parameters}.}
\label{fig:RKcontour}
\end{figure}

In \fig{fig:RKcontour} we display the contour lines  $R_K$ and $R_{K^*}$ in the $m_{R'_1}$-$\tan\beta$ plane.
The shaded regions indicate the 1-$\sigma$ regions consistent with present data. The increase of $\tan\beta$ with
increasing $m_{R_{1}'}$ can easily be understood from \eqt{eq:C9R2}{eq:C10R2} 
as the new physics contributions
scale as 
\begin{align}
\frac{Y^2_4}{m^2_{R'_1}} \propto \frac{\tan^2\beta}{m^2_{R'_1}}\,.
\end{align}  

\begin{figure}[t]
\includegraphics[width=\linewidth]{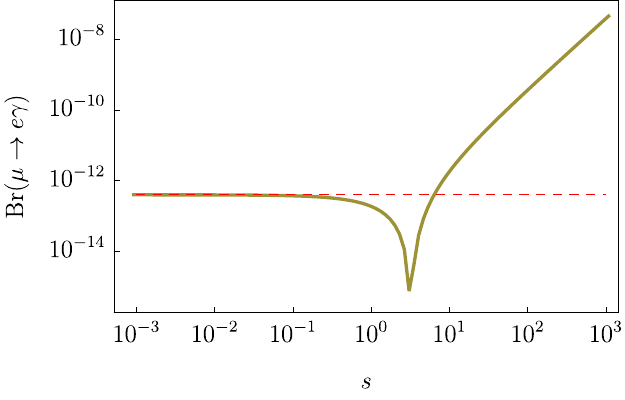}
\caption{BR($\mu\to e  \gamma$) as a function of $s$ which rescales the values of $Y_2$ in
\tab{tab:parameters}. The other parameters are fixed as before.
}
\label{fig:BRmuegamma}
\end{figure}

\begin{figure}[t]
\includegraphics[width=\linewidth]{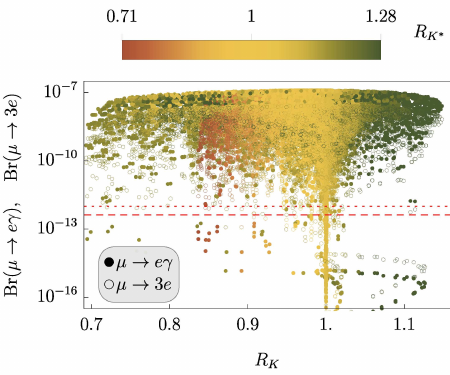}
\caption{Correlation between BR($\mu\to e\gamma$), BR($\mu\to 3 e$)  and
$R_K$ obtained by varying $\phi_{12},\phi_{13},\phi_{23}$ in equal steps of $\pi/27$ in the range $[0,\pi]$ each.  The other parameters are fixed as given in \tab{tab:parameters}. 
The color code indicates the value of $R_{K^*}$.
} 
\label{fig:BRmu3e}
\end{figure}

We recall that the leptoquarks in general also contribute to lepton-flavor violating decays
of the muon such as  \mbox{$\mu\to e\gamma$}, see for instance \cite{Crivellin:2017dsk}.
There are two main contributions to this observable in the current setting coming, namely, from heavy-neutrino and $W$-boson loops as
well as leptoquark and quark loops. As an example, in \fig{fig:BRmuegamma} we show $\text{BR}(\mu\rightarrow e \gamma)$ as a function of an extra factor $s$ rescaling the $Y_{2}$ eigenvalues in \tab{tab:parameters} into $Y_2\to s Y_{2}$. For $s\gsim 5$ the
leptoquark loops dominate whereas for $s\le 1$ the neutrino loops are dominant. The narrow minimum is due
to a negative interference between both contributions\footnote{In principle, $s$ (and, hence, $Y_{2}$) may be even larger than that indicated in~\fig{fig:BRmuegamma} if off-diagonal elements of $Y_5$ were invoked together with this negative interference. 
However, as
$Y_2$ does not enter the calculation of $R_{K^{(*)}}$ at the lowest order, we do not investigate this further.}.

Let us point out that in the current model the bounds from $\mu \to 3 e$ are 
in general stronger than those from
\mbox{$\mu\to e  \gamma$} in the range interesting for $R_K$ and
$R_{K^*}$, see \fig{fig:BRmu3e}.  
 The main reason for this is the  negative
interference in $\mu\to e\gamma$ discussed above which does neither take place in the $Z$-penguins nor in the box-contributions to $\mu\rightarrow 3 e$ (see Fig. \ref{fig:DiaMu3e})  for the same set of parameters.  
\begin{figure}[t]
\includegraphics[width=\linewidth]{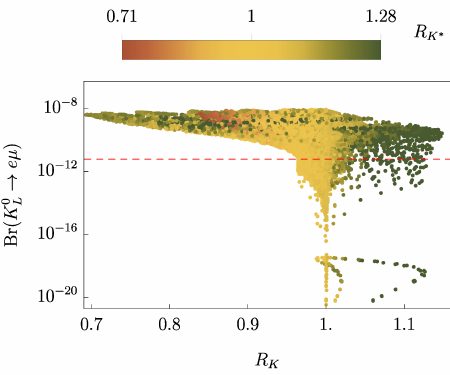}
\caption{Correlation between BR($K_L\to e \mu$) and $R_K$ obtained by varying $\phi_{ij}$ as in \fig{fig:BRmu3e}. The other parameters are fixed as given in \tab{tab:parameters}.
The color code indicates 
the value of $R_{K^*}$ for each point.
} 
\label{fig:BRKLemu}
\end{figure}

In addition, we have checked that rare $\tau$ decays do not impose any         constraints in the   $R_{K^{(*)}}$-interesting regions. 
The same holds for rare
$b$-decays, such as $B_{s,d} \to \mu^+\mu^-$, $B\to s \gamma$, $B\to X_s \nu\nu$.
Taking only the couplings of the scalar sector given in ref.~\cite{Perez:2013osa} would yield \mbox{$m^2_{G^{+,0}}=m^2_{R'_1} + O(v^2_{\rm ew})$} leading to too large contributions to $\Delta M_K$ and 
$\Delta M_B$. However, this relationship gets broken by $\lambda_{14}$ in \eq{eq:Vscalar}
implying that also these bounds can be avoided. 

In any case, there is another stringent constraint to be considered, namely
the bound on $K_L \to e \mu$. This mode is
usually used as a limit on the mass of vector leptoquarks but
it is typically not being taken into account for the scalar ones. As can be seen in~\fig{fig:BRKLemu}, in the interesting
region for  $R_K$ the bound is violated by several orders of magnitude, thus ruling out
this model even in those tuned parts of the parameter space where all other constraints can be satisfied.

It has been argued in \cite{Calibbi:2017qbu} that additional fermions in
vector-like representations of the gauge group can reduce the couplings of the
SM-fermions to vector leptoquarks which, in turn, may be used for lowering the generic experimental limits for their masses. The same mechanism can in principle work also for the couplings of the scalar leptoquarks such that the constraints due to
$K_L\to e \mu$ can be satisfied. However, a detailed exploration of this aspect is
beyond the scope of this paper and will be elaborated on elsewhere \cite{longpaper}.

We note for completeness that leptoquarks with masses of about 1 TeV are already constrained by the LHC searches. These, however, typically focus on the situation when the decays are dominated by one channel; in the scenarios where the LQs interact through multiple couplings the corresponding bounds must be re-evaluated. 


We would also like to point out that the ``LQ beta-decay modes'', i.e., the decays of a heavier LQ into its lighter $SU(2)_{L}$ companion and $W$ have not been considered so far within the collider searches. Nevertheless, depending on the exact mass splitting, they may be of significant interest, especially if the on-shell $W$ production is kinematically allowed\footnote{Let us note that in the model of our interest the LQ-doublet mass splitting would
be below $m_W$ if we were to consider only the potential given in \cite{Perez:2013osa} but can be larger
once the additional terms as in \eq{eq:Vscalar} are included.}. 



\section{Discussion and conclusions}
\label{sec:conclusions}

Motived by the successful attempts to explain the observed values of $R_{D^{(*)}}$ and $R_{K^{(*)}}$
by leptoquarks we study a unified $SU(4)_\mathrm C\otimes SU(2)_\mathrm L\otimes U(1)_\mathrm R$ gauge model in order to demonstrate the challenges one faces when all additional relations inherent to a unified scenario are taken into account. Among these, the dominant role is typically played by the constraints on the Yukawa couplings from the quark and lepton masses and mixing data and/or the tight connection between the relevant gauge coupling and~$\alpha_{s}$.

The model under consideration contains three different types of leptoquarks: a hypercharge-2/3 vector leptoquark and a pair of
scalar leptoquarks with hypercharges $1/6$ and $7/6$.  In its minimal version, with the SM fermion sector extended such that it supports the inverse-seesaw mechanism for neutrinos,
one finds~\cite{Perez:2013osa} that the kaon physics constrains the mass of the vector leptoquark
to such an extent that it cannot significantly impact the $B$-physics observables.

 We have shown that in the setting under consideration one can get the hypercharge-$7/6$ scalar leptoquark in
the TeV range while, at the same time, have automatically the hypercharge-$1/6$ scalar leptoquark
rather heavy. 
With this scalar sector, the low-energy effective operator structure of the model  at the scale of $B$-mesons is such that it can accommodate $R_{K}$ and $R_{K^{*}}$ but neither $R_{D}$  nor $R_{D^{*}}$.

We find that the allowed parameter space for $R_{K}$ and $R_{K^{*}}$ gets severely constrained
by the bounds on rare muon decays; in particular, $\mu\to 3 e$ is more
important than  $\mu\to e\gamma$. Neither lepton flavor violating $\tau$-decays nor
other $B$-physics observables lead to additional constraints. However, it turns
out that no points in the available parameter space are compatible with $K_L\to e \mu$.

On the other hand, this does not imply that this kind of a leptoquark model is ruled out straight away as
an explanation of $R_{K^{(*)}}$ because one can always enlarge
the fermion sector by vector-like representations. In this way one may in principle
reduce the couplings to the muon by mixing effects and, thus, avoid the
bound due to $K_L\to e \mu$. This goes beyond the scope of this letter and will be elaborated on in a future study.

\section*{Acknowledgements}
We thank H.~Kole\v{s}ov\'a for discussion in the initial phase of this project.
T.F., P.M.\ and W.P.\ have been supported by  the DFG, project nr.\ PO-1337/7-1.
FS is supported by the ERC Recognition Award ERC-RA-0008 of the Helmholtz Association.
M.H. and M.M. acknowledge the support from the Grant Agency of the Czech Republic, Project No. 17-04902S.

\appendix

\section{Baryon number conservation}
\label{sec:proton_decay}

It has been noted in \cite{Perez:2013osa} that the model has an approximate extra $U(1)_\mathrm F$ symmetry corresponding to 
the fermion number $F$, which is explicitly broken by the Majorana mass term for $N$. However, there is another independent 
accidental global symmetry $U(1)_\mathrm M$, the charges of which are
\begin{align}
\begin{split}
M({F_L}^\alpha)=M(\chi^\alpha)&=+1,
\\
M({f^c_{u}}_\alpha)=M({f^c_d}_\alpha)&=-1,
\\
M(\Phi^\alpha_{\,\beta})=M(H)=M(N)&=0,
\\
M({A_\mu}^\alpha_{\, \beta}) = M(W_\mu) = M(B_\mu) &=0.
\end{split}\end{align}
Here $\alpha, \beta$ denote $SU(4)$ vector-like indices. 
Notice that 
we can obtain the $M$-charges of all the field multiplets by the prescription
\begin{align}
M = \big(\#\,\text{upper}\;SU(4)\; \text{indices}\big) -\big( \#\,\text{lower}\;SU(4)\; \text{indices}\big).
\label{eq:inductionRule}
\end{align}

The $U(1)_\mathrm M$ symmetry of each term in the Lagrangian is then guaranteed by the fact that 
every upper $SU(4)$ index is contracted to a lower one, all carried by the dynamical fields. It 
is also clear that any hypothetical $M$--violating but $SU(4)$--preserving term necessarily 
contains the antisymmetric tensor $\varepsilon_{\alpha\beta\gamma\delta}$. Hence, such type of a 
symmetry is realized in any $SU(4)$ model whose field content does not allow for the 
$SU(4)$ Levi-Civita symbol to occur in the interaction Lagrangian at the renormalizable level; 
for example, in \cite{Foot:1997pb}, the corresponding number is called $B'$. 

Having the $M$-charge at hand, we can combine it with the gauge charge~(\ref{T15}) as
\begin{align}\label{eq:defOfBaryonNumber}
B &= \frac{1}{4}\left(M + \left[B \! - \! L\right]\right),
\end{align}
which obviously yields the baryon number (see \tab{tab:U1charges}). 

As one can verify readily, in the model under consideration both $[B\!-\!L]$ and $M$ are spontaneously broken by~$\langle \chi\rangle$ whilst their sum~(\ref{eq:defOfBaryonNumber}) remains a good symmetry even in the asymmetric 
phase. 
%

On a more general ground, one can rephrase the same argument as follows: 
If $M$ defined as~(\ref{eq:inductionRule}) is a good symmetry of the unbroken-phase theory, there is no $SU(4)_{\rm C}$-Levi-Civita tensor in its Lagrangian ${\cal L}_{4}$.  This means that there is no $SU(3)_{\rm c}$-Levi-Civita tensor in the broken-phase Lagrangian ${\cal L}_{3}$ either. Consequently, a global charge defined as
\begin{align}
\begin{split}
3B &= (\#\text{ upper }SU(3)_{\rm c}\text{ indices}) 
\\&- (\#\text{ lower }SU(3)_{\rm c}\text{ indices})
\end{split}
\label{eq:inductionRuleB}
\end{align}
generates a good symmetry of ${\cal L}_{3}$ and, hence, $B$ -- the usual SM baryon number -- is perturbatively conserved. 

Let us also note that there are two slightly different candidates for the lepton number, none of which is, however, related to a fully conserved quantity in our model. 
The first option is intuitive,
\begin{align}
L =B - \left[B\!-\!L\right].
\end{align}
Here, $U(1)_\mathrm L$ is a good symmetry of the classical action but it is spontaneously broken, together with $[B\!-\!L]$, by~$\langle\chi\rangle$. The alternative,
\begin{align}
L'= F - 3B,
\end{align}
is, on the other hand, preserved by the vacuum but explicitly broken by the Majorana mass term because $F$ is so.

\bibliographystyle{h-physrev5}
\bibliography{LQ}

\end{document}